\documentclass[10pt,aps,onecolumn,floatfix,superscriptaddress,longbibliography]{revtex4} 
\usepackage{graphicx,amsmath}
\usepackage{color}

\renewcommand{\Re}{\operatorname{Re}}

\newcommand{\St}{\operatorname{St}}

\begin{document}

\title{Self-similar decay of high Reynolds number Taylor-Couette turbulence}
\author{Ruben A. Verschoof} 
\author{Sander G. Huisman}  
\author{Roeland C.A. van der Veen}  
\affiliation{Department of Applied Physics, MESA+ institute and J. M. Burgers Center for Fluid Dynamics, University of Twente, P.O. Box 217, 7500 AE Enschede, The Netherlands}

\author{Chao Sun}
\email{chaosun@tsinghua.edu.ch}
\affiliation{Center for Combustion Energy and Department of Thermal Engineering, Tsinghua University, 100084 Beijing, China}

\affiliation{Department of Applied Physics and J. M. Burgers Center for Fluid Dynamics, University of Twente, P.O. Box 217, 7500 AE Enschede, The Netherlands}

\author{Detlef Lohse}
\email{d.lohse@utwente.nl}
\affiliation{Department of Applied Physics and J. M. Burgers Center for Fluid Dynamics, University of Twente, P.O. Box 217, 7500 AE Enschede, The Netherlands}
\affiliation{Max Planck Institute for Dynamics and Self-Organisation, 37077 G\"{o}ttingen, Germany}

\date{\today}

\begin{abstract} 
We study the decay of high-Reynolds number Taylor-Couette turbulence, i.e. the turbulent flow between two 
coaxial rotating cylinders. 
To do so, the rotation of the inner cylinder ($\Re_i=2 \times 10^6$, the outer cylinder is at rest) is stopped within 12 s, thus fully removing the energy input to the system. 
Using a combination of laser Doppler anemometry and particle image velocimetry measurements, 
six   decay decades of the kinetic  energy could be   captured.
 First, in the absence of cylinder rotation, the flow-velocity during the decay does not develop any
  height dependence in contrast to the well-known Taylor vortex state. 
  Second, the radial profile of the azimuthal velocity is found to be self-similar. Nonetheless, 
  the decay of this wall-bounded inhomogeneous turbulent flow does not follow a strict power law as 
  for decaying turbulent homogeneous isotropic flows,   but it is faster, due to the strong viscous drag applied by the bounding walls.
   We theoretically describe the decay in a quantitative way by taking the effects of additional friction at the walls into account.

\end{abstract}

\maketitle

Turbulence is a phenomenon  far from equilibrium: Turbulent flow is driven in one or the other way
by some energy input and at the same time energy is dissipated, predominantly  
(but not exclusively) at the smaller scales. 
For statistically stationary turbulence, this balance is reflected in the famous picture of the 
Richardson-Kolmogorov energy cascade \cite{kol41a,pop00}. While the driving on large scales 
clearly is non-universal, depending on the flow geometry and stirring mechanism, the energy dissipation
mechanism has been hypothesized to be self-similar  \cite{kol41b,kol41c,saf67,sre84,Vassilicos2015,Sinhuber2014}. 

How exactly is the energy taken out of the system? 
A good way to find out is to turn off the driving
and follow the then  decaying turbulence, as then all scales are probed during the decay process.
This has been done in various studies over the last
decades for homogeneous isotropic turbulence (HIT).
Experimentally, the focus of attention was 
on  grid-induced turbulence 
\cite{George1992,smi93a,George2009,sta99,Skrbek2000,Burattini2005,Thormann2014,Sinhuber2014},
whereas in numerical simulations periodic boundary conditions were used \cite{Biferale2003,Burattini2006,ish06,Teitelbaum2009}. 
To what degree the decay of the turbulence depends on the initial conditions 
\cite{lav07,val11,hur07} and 
whether or not it is  self-similar   
has controversially been debated  \cite{saf67,ant03,Biferale2003,eyi00,George2009,mel11,Riboux2010,dav11}. 
We note that for HIT,  already from 
dimensional analysis  one obtains power laws for the temporal evolution of the vorticity and kinetic energy
in decaying turbulence, namely $\omega(t) \propto t^{-3/2}$ 
 and $k(t) \propto t^{-2}$, respectively, in good agreement with many measurements \cite{smi93a, ll87, sta99}.
 These scaling laws are also obtained
\cite{loh94a} when employing the `variable range mean field theory' of Ref.\ \cite{eff87}, developed for HIT. In that way, the late-time behavior, when the flow is already viscosity dominated, can also be calculated, allowing for the  calculation of the lifetime of the decaying turbulence \cite{loh94a}.

However, real turbulence is neither homogeneous nor
isotropic, but it  has anisotropies and is 
wall-bounded,
with a considerable fraction of the dissipation taking place in the corresponding boundary layers. 
Studies on stedecay of fully developed turbulence flow 
 in wall-bounded flows are however scarce \cite{Touil2002}, though exploring 
 the decay of such flows would teach us about the energy dissipation in the boundary layers and its 
 possible universality. 
 The reason for the scarcity of such studies may be that for the most canonical and best-studied wall-bounded
 flow, namely, pipe flow \cite{smi11,hul12,mar10b,smi13}, the decaying turbulent 
 flow is flushed away downstream so that it is hard to study it.

\begin{figure}[!htbp]
\centering
\includegraphics[scale=0.75]{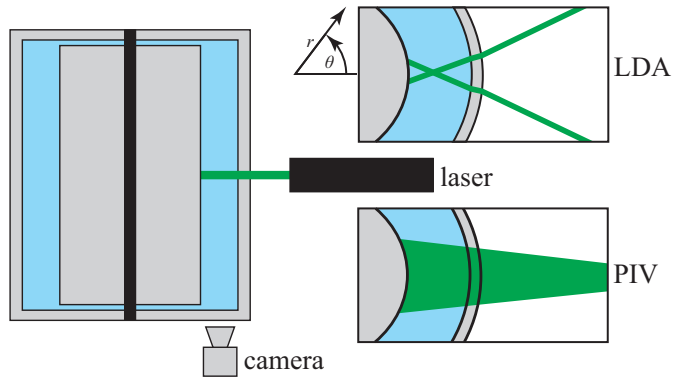}
\caption{Schematic of the vertical cross-section of the T$^3$C facility. The laser beams are in the horizontal plane ($r,\theta$) at midheight; $z = L/2$ (unless stated otherwise) for both the LDA and PIV measurements. The top right inset shows the horizontal cross-section, showing the LDA beams (not to scale). The beams refract twice on the OC and intersect at the middle of the gap ($r = r_m$), giving the local velocity component $u_{\theta}$. The bottom right inset shows, for the PIV measurements, particles are illuminated by a thin laser sheet. We use the viewing windows in the end plate to look at the flow from the bottom, thus obtaining the velocity components $u_{\theta}$ and $u_r$ in the ($r,\theta$) plane.}
\label{fig:fig1}
\end{figure}

This problem is avoided in confined  and at the same time 
  closed turbulent flows, such as Rayleigh-B\'enard flow \cite{ahl09,loh10} or
Taylor-Couette (TC) flow \cite{don91,far14,gro16}, 
 i.e.,  the flow 
 between two independently rotating co-axial cylinders (Fig.\ \ref{fig:fig1}). 
 Indeed, turbulent TC flow is neither homogeneous nor isotropic, due to coherent structures that persist also at high Reynolds numbers \cite{hui14, lew99}, and the boundary layers play 
 the determining  role in the angular momentum transfer from the inner to the outer cylinder \cite{hui13,ost14pof}.
 
In this study, we employ the TC system to study the temporal and spatial behavior
 of decaying confined and wall-bounded 
 turbulence, 
 and compare it with the known results for HIT, thus complementing the study of Ref.\ \cite{Sinhuber2014}
 for decaying homogeneous isotropic turbulence. 
 We suddenly stop the inner cylinder rotation (similarly as in Ref.\ \cite{bor10},
 which focused on the decay of turbulent puffs for much
 lower Reynolds number) 
 and then measure the velocity field over time. 
 We find that the decay models for HIT  \cite{loh94a,Sinhuber2014} are
  insufficient to describe the data, but when extending them
   by explicitly taking the wall friction into consideration, the measured data can be well described. 
 Though the decay does not follow a power law due to the wall friction, 
 the velocity profiles are still self-similar. 
In the study we restrict ourselves to fixed outer cylinder and decaying
 flow; for a numerical study on flow stabilization by a corotating outer cylinder we refer the reader to Ref.\ \cite{ost14jfmr}.

The experiments were performed at the Twente Turbulent Taylor-Couette facility (T$^3$C) \cite{gil11a}, consisting of two independently rotating concentric smooth cylinders. The setup has an inner cylinder (IC) with a radius of $r_i = 200$ mm and an outer cylinder (OC) with a radius of $r_o = 279$ mm, giving a mean radius $r_m=(r_i+r_o)/2=239.5$ mm, a radius ratio of $\eta = r_i/r_o = 0.716$ and a gap width $d=r_o-r_i= 79$ mm. The IC can rotate up to $f_i=20$ Hz, resulting in a Reynolds number up to $\text{Re}_i = 2\pi f_i r_i d/\nu = 2 \times 10^6$ with water as the  working fluid at $T =20^{\circ}$C. The cylinders have a height of $L = 927$ mm, giving an aspect ratio of $\Gamma = L/d = 11.7$. The transparent acrylic OC allows for non-intrusive optical measurements. The end plates, which are partly transparent, are fixed to the OC. The velocity is measured using two non-intrusive optical methods: particle image velocimetry (PIV) and laser Doppler anemometry (LDA), as shown in Fig.\ \ref{fig:fig1}. The LDA measurements give the azimuthal velocity $u_{\theta}(t)$ at mid-gap ($r=r_m$) and at several heights. The water is seeded with 5$\mu$m diameter polyamide tracer particles with a density of 1.03 g/cm$^3$.  The laser beams are focused in the middle of the gap, i.e.\ at $r = r_m$. Using numerical ray-tracing, the curvature effects of the OC are accounted for \cite{Huisman2012a}. The PIV measurements are performed in the $\theta-r$ plane at mid-height ($z=L/2$), using a high-resolution camera \footnote{pco, pco.edge camera, double frame sCMOS, 2560 pixel $\times 2160$ pixel resolution, operated in dual frame mode.}, operating at 20 Hz. The spatial resolution of the PIV measurements is 0.04 mm/pixel, with interrogation windows of 32 pixel $\times$ 32 pixel. The flow was illuminated from the side with a pulsed Nd:YLF laser \footnote{Litron, LDY303HE Series, dual-cavity, pulsed Nd:YLF PIV Laser System. The sheet thickness was approximately 1 mm.}, with which a horizontal light sheet is created (fig. \ref{fig:fig1}). The water was seeded with 20 $\mu$m polyamide tracer particles. Because of the large velocity range of our measurements, several measurements with a changing $\Delta t$ are performed ($50~\mu \text{s} \leq \Delta t \leq 50$ ms), so that the entire velocity range is fully captured. The PIV measurements were processed to give both the radial velocity $u_r(\theta,r,t)$ and azimuthal velocity $u_{\theta}(\theta,r,t)$. The Stokes number of the seeding particles are always smaller than $\St = \tau_p/\tau_{\eta} < 0.2$, so the particles faithfully follow the flow  \cite{pop00, Mei1996}.

We first drive the turbulence at a rotation rate of  $f_i = 20$ Hz ($\Re_i=2\times 10^6$) of the IC, while the OC is at rest, allowing for the development of a statistically stationary state. We then decelerated the IC within approximately 12 s linearly down to $f_i = 0$ Hz, so, starting from $t = 0$ s, there is zero energy input. The deceleration rate is limited by the braking power of the electric motor. The deceleration time is much smaller than the typical time scale for turbulence decay ($\tau = d^2/\nu \approx 6 \times 10^3$ s). The velocity measurements start when the IC has come to rest, so at $t=0$,  $f_i = f_o = 0$ Hz. 

\begin{figure}[!htbp]
\begin{center}
\includegraphics{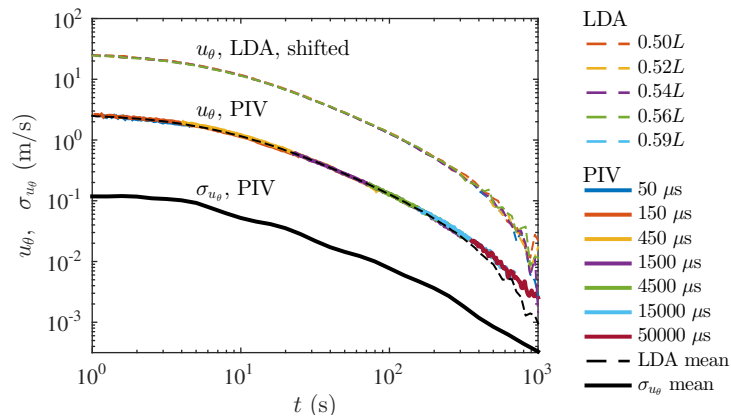}
\caption{Midgap azimuthal velocity as a function of time at mid-height. PIV measurements with seven different interframe times $\Delta t$ were performed, as shown in the legend, to produce accurate results over the entire velocity range. The PIV measurements are averaged azimuthally and radially; we average over $r_m-4~\text{mm} <r <r_m+4$ mm, corresponding to 10\% of the gap width. These results are confirmed by LDA measurements performed at several heights (dashed lines), which are shifted by one decade for clarity. The standard deviation $\sigma_{u_\theta}$, a measure for the spatial velocity fluctuations,
 is shown as the solid black line. The data are averaged azimuthally and radially as described above, and binned using logarithmic bins of 0.1 decades. The measurements cover three orders of magnitude of the velocity, corresponding to six orders of magnitude in kinetic energy. The measurement uncertainty roughly corresponds to the width of the lines.}
\label{fig:fig2}
\end{center}
\end{figure}

In Fig.\ \ref{fig:fig2}, the azimuthal velocity decay 
$u_\theta(t)$  is shown. The results obtained with PIV and LDA are the same;
 the LDA results only start to deviate from the PIV measurements when the measured velocities are close to the dynamic range of the LDA system.  Viscous friction dissipates the energy, bringing the fluid eventually to rest. 
 Also the spatial velocity fluctuations, 
  characterized by the standard deviation of the azimuthal velocity fluctuations $\sigma_{u_\theta}(t)$ decay
   in a 
   very similar way, see Figs.\ \ref{fig:fig2} and 
 \ref{fig:fig4}.
 The LDA data (dashed lines shown in fig.\ \ref{fig:fig2}) 
  are measured at different heights using LDA; their collapse indicates that during the decay no Taylor rolls develop, which would lead to a height-dependence of the profiles. This is in contrast to TC flow with increasing inner cylinder rotation, where with increasing $\Re_i$ first Taylor rolls develop \cite{and86}, before one arrives at the structureless fully developed turbulent state (for the chosen geometry) \cite{lat92a,gro16}. The reason for this difference is that in the constantly rotating case angular momentum is transported from the inner to the outer cylinder, whereas in the decaying case the angular momentum is transported from the bulk to both walls, i.e., a net momentum transport between the cylinders is absent.
  From earlier work \cite{hui13}, we know that the normalized velocity profiles in the bulk are nearly Re independent and height independent over a large range of Reynolds numbers. Here we focus on the bulk flow velocity where the Re-independence holds. Correspondingly, we would get similar results for different Re-measurements in the turbulent regime. The axial and radial velocities are approximately 50-100 times smaller than the azimuthal velocity, so their contributions to the total kinetic energy are negligible.


\begin{figure}[!htbp]
\begin{center}
\includegraphics{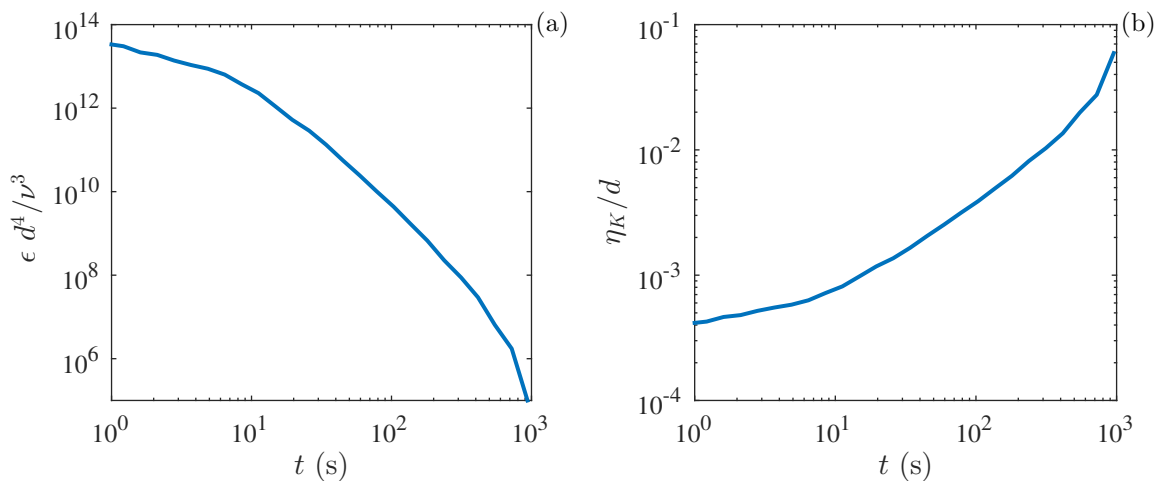}
\caption{
(a) 
Energy dissipation rate, normalized with $\nu^3/d^4$, as a function of time. Here $\epsilon$ is calculated from the PIV data as shown in Fig.\ \ref{fig:fig2}. $\epsilon$ drops by more than 8 orders of magnitude.
(b) Kolmogorov length scale $\eta_K$ as a function of time, normalized with the gap width $d$. As time progresses, the separation of scales becomes smaller, although $\eta_K$ remains small. }
\label{fig:fig3}
\end{center}
\end{figure}

Turbulence is characterized by a fluid motion over a large range of length scales. In TC flow, the upper limit is the gap width $d$ and the smallest length scale is the Kolmogorov scale $\eta_K$, which is defined as $\eta_K = \left( \nu^3/\epsilon \right)^{1/4} $.
From the bulk velocity as measured with PIV, we calculate the energy dissipation rate from the change in velocity over time, i.e. $\epsilon = d(\frac{1}{2} u_{\theta}^2)/dt$. As shown in Fig.\ \ref{fig:fig3}, as the velocity decreases, also the energy dissipation rate becomes smaller. Clearly, the dissipative length scale changes over time, though $\eta_K$ only remains a fraction of the gap. Consequently, we cannot faithfully resolve spatial gradients in the flow with our PIV data.

\begin{figure}[!htbp]
\begin{center}
\includegraphics{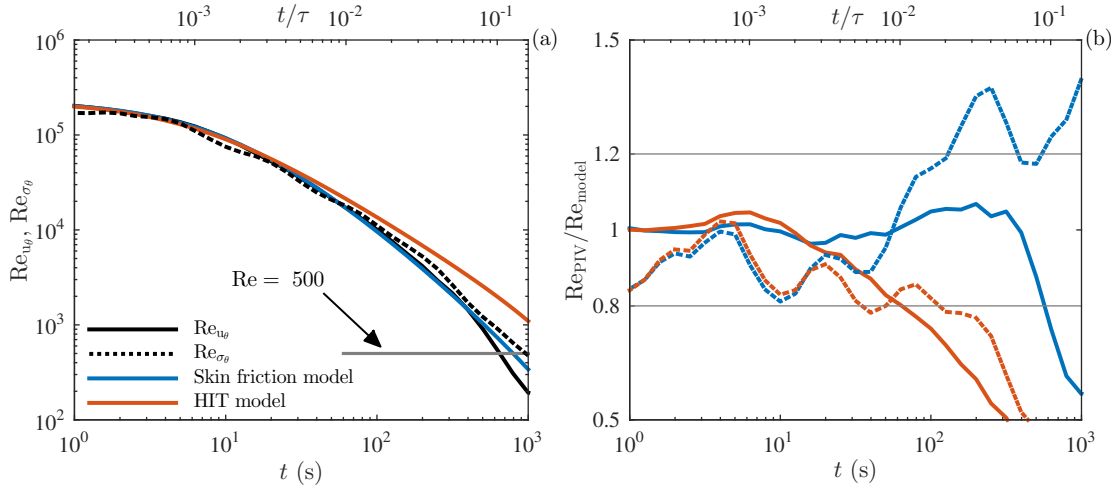}
\caption{(a) 
The PIV results for $\Re_{u_\theta}$ and $\Re_{\sigma_\theta}$, 
binned using logarithmic bins of 0.1 decades: 
The scale on the y-axis refers to  $\Re_{u_\theta}$, whereas the one for $\Re_{\sigma_\theta}$ is vertically shifted to show that the decay of the spatial fluctuations and the mean is the same. On the horizontal axis, both the real time and the non-dimensionalized time (normalized by $\tau = d^2/\nu$) are shown. 
Included in the graph are the results for the HIT model 
of Ref.\ \cite{loh94a} and those of the Prandtl-von K\'arm\'an 
skin friction model (eq.\  (2)), which includes the effects of the walls. Below the short line at $\Re = 500$, thermal effects set in. 
(b) Ratio between the measurements and respectively 
the HIT model (solid red line) and the skin friction model (solid blue line). 
 The respective dashed lines show the ratio between the fluctuation decay and the two models. }
\label{fig:fig4}
\end{center}
\end{figure}

To compare the experimental data on the decay of the velocity and their fluctuations with theory, 
we first define the respective Reynolds numbers, namely $\Re_{u_\theta}(t)  = u_{\theta}(r=r_m,t)d/\nu$
taken at mid-gap $r_m$ 
and  $\Re_{\sigma_\theta}(t)  = const \times \sigma_{u_\theta}(r=r_m,t)d/\nu$ for the fluctuations, which we
have rescaled with a constant so that it collapses with $\Re_{u_\theta}$ at $t=0$, i.e.\ $const = \Re_{u_{\theta}}(0)/\left(\sigma_{u_\theta}(0)d/\nu  \right)$. The curves
show
that the decay  of  the velocity itself and the fluctuations is the same [see Fig.\ \ref{fig:fig4}a]. 
We then compare the decay of $\Re(t)$ with the one predicted for the theory of 
HIT, as it follows from a numerical integration of 
the ordinary differential equation obtained in the model of Ref.\ \cite{loh94a},
\begin{align}
\dot{\Re} &=-\frac{1}{3} \frac{\nu}{d^2}c_{\mathrm{HIT}}(\Re) \Re^2,
\label{eq:Re}
\end{align}
with $c_{\mathrm{HIT}} (\Re)$ given by Eq.\ (6) of \cite{loh94a}. From Fig.\ \ref{fig:fig4} we see that, though in the beginning
the decay is reasonably well described, at a later time the decay experimentally 
found in this wall-bounded flow is much faster
than resulting from the  model for HIT.

We therefore replace the model for $c_{\mathrm{HIT}} (\Re)$ in Eq.\ (\ref{eq:Re}) by a model  
  for  {\it wall-bounded}  flow, 
 namely by a friction factor $c_f(\Re)$ following from 
 the Prandtl-von K\'arm\'an skin friction law \cite{pop00, lew99,smi11},
 \begin{align}
\begin{split}
\frac{1}{\sqrt{c_f}}&= a \log_{10}(\Re \sqrt{c_f}) + b.
\label{eq:cf}
\end{split}
\end{align}

For pipe flow, a good description of various 
experimental data can be achieved with $a =1.9$ and $b =-0.3$ \cite{Zagarola1998}. These values depend on the boundary conditions of the flow, i.e. on the geometry and whether or not the flow is actively driven or decaying. For decaying turbulence in the TC geometry we find that a good least-square fit 
 of this model to the decay of  $\Re(t)$ is achieved with $a= 2.72$ and $b=-2.22$ (see  Fig.\ \ref{fig:fig4}).  As can be seen, due to the extra friction in the wall regions the decay is now 
faster than the decay observed in HIT and much better and longer agrees with the experimental data,
reflecting that 
 the no-slip boundary conditions force the fluid to slow down faster.

How long do the respective models for HIT \cite{loh94a} and the Prandtl-van Karman skin
 friction hold?
 We define the beginning of the discrepancy between data and models
 to be $|1- \Re_\text{PIV}/\Re_\text{model} |=0.2 $, which is visible as thin gray lines in Fig.\ \ref{fig:fig4}(b). From this definition, we calculate that the discrepancy between the $\Re_{u_\theta}$ and the HIT model starts at $t=45$ s and the one
  between $\Re_{u_\theta}$ and the friction model at $t=530$ s.

 \begin{figure}[htbp]
\begin{center}
\includegraphics{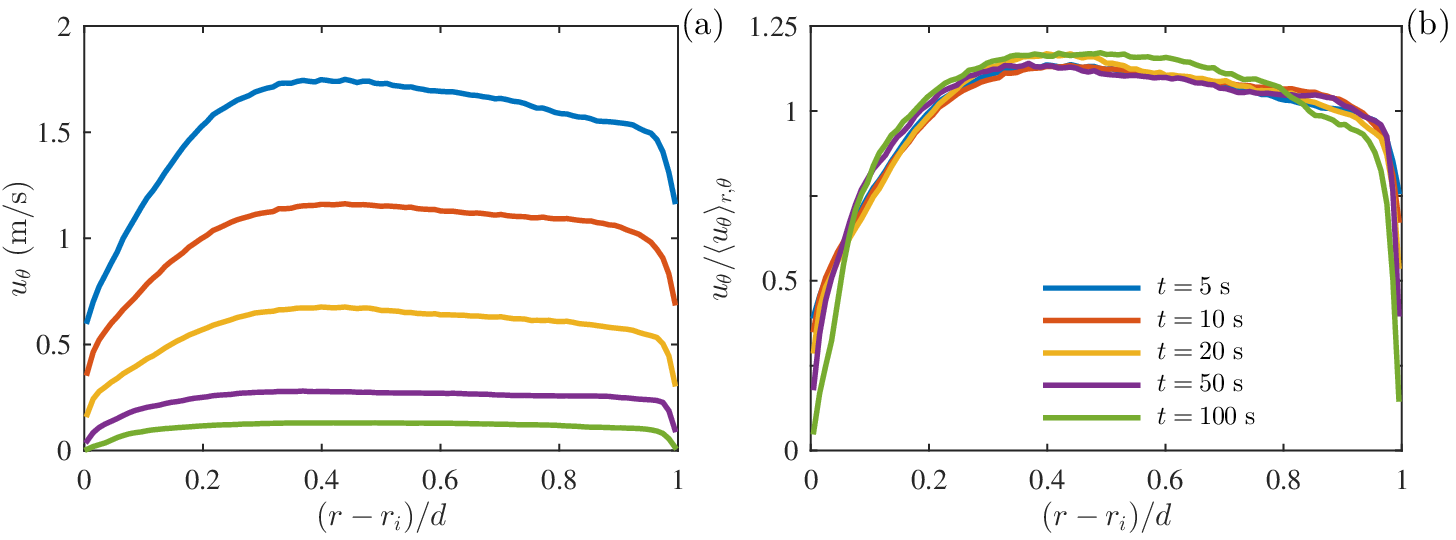}
\caption{Azimuthal velocity profiles $ u_{\theta}(r)$ for $z=L/2$, averaged over $\theta$. The decelerating effects of the walls 
(left and right edges of the figure) can be seen clearly. (b) The velocity is normalized with the (spatial) mean azimuthal velocity, $\langle u_{\theta} \rangle_{r,\theta}(t)$. The normalized velocity profiles overlap, indicating the self-similarity of the velocity profile during the decay.}
\label{fig:fig5}

\includegraphics[scale=1.2]{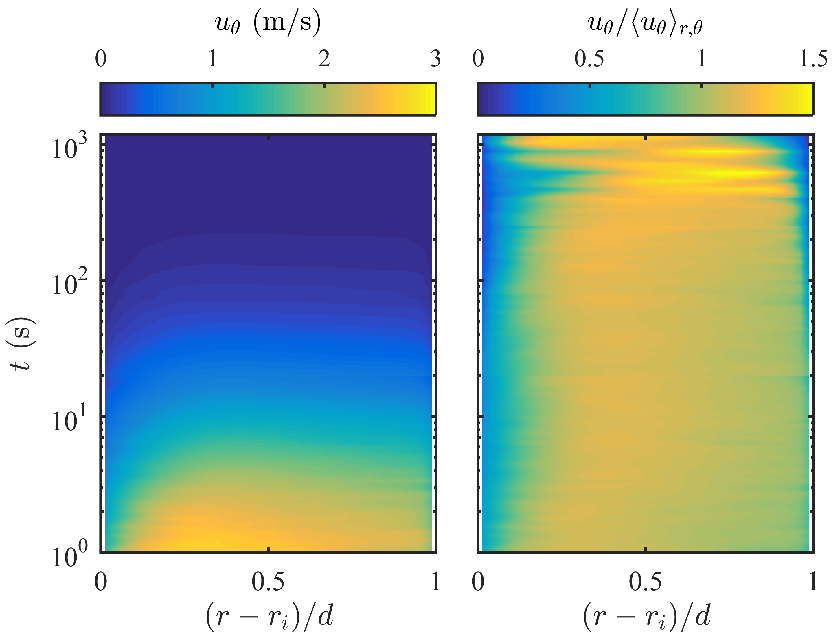}
\caption{(a) Measured velocity $u_{\theta}$ as a function of time, resulting from seven  PIV measurements with changing $\Delta t$ and averaged over $\theta$.
(b) Same results as shown in (a) but now normalized with the mean velocity $\langle u_{\theta}(t) \rangle_{r,\theta}$. The normalized velocity is self-similar up to $t\approx 400$ s.}
\label{fig:fig6}
\end{center}
\end{figure}

As was discussed in the Introduction, self-similarity is commonly assumed and observed \cite{saf67,George1992}  in the decay of HIT flows. The question is whether self-similar flow fields still exist in the decay of inhomogeneous wall-bounded
 turbulence with a strong shear. By analyzing several instantaneous velocity profiles (see fig.\ 
 \ref{fig:fig5}), we found that also for this inhomogeneous turbulence the normalized velocity profiles are self-similar during the decay, see Fig.\ \ref{fig:fig6}. We find that the normalized velocity profile is self-similar up to $t \approx 400$ s. Hitherto, a self-similar decay has not yet been observed for wall-bounded inhomogeneous turbulence, and it is remarkable 
that also in this highly inhomogeneous and anisotropic flow a self-similar decay exists. Eventually the self-similarity breaks down, possibly due to thermal convection. Residual cooling in the end plates causes small temperature differences and thus thermal convection is estimated to start from $\Re \approx 500$ (see Ref.~\cite{gil11a} for a detailed discussion). Therefore, the results after $t \approx 600$ s are dominated by effects other than the initial velocity and the decay process. As can be seen in Fig.~\ref{fig:fig4}, this roughly coincides with the moment the model starts to deviate from our measurements.

In conclusion, we measured six decades of the decaying energy in Taylor-Couette flow after the cylinders were halted. During the decay, no height dependence of the flow develops, which is in contrast to the upstarting case, in which the well-known Taylor vortices develop. The azimuthal velocity profile was found to be self-similar. Nonetheless, the kinetic energy in this wall-bounded flow decays faster than observed for homogeneous isotropic turbulent flows. This accelerated decay is due to the additional friction with the walls. We successfully modeled this accelerated decay by using a friction coefficient in which the 
Prandtl-von K\'arm\'an skin friction law for wall-bounded flow 
 is used to model  $c_f(\Re)$. With this model, both the decay of the mean and the fluctuations could be described successfully. We hope that this work will stimulate further investigations into the decay of wall-bounded (and thus non- sotropic and inhomogeneous) turbulence in other flow geometries, disentangling the universal and non universal features.

\begin{acknowledgments}
The authors gratefully acknowledge informative discussions with and the valuable input of Bruno Eckhardt. We would also like to thank Gert-Wim Bruggert and Martin Bos for their continual technical support over the years, and Leen van \hyphenation{Wijn-gaar-den}Wijngaarden, Dennis Bakhuis, and Rodrigo Ezeta Aparicio for various stimulating discussions. This work was supported by the Dutch Foundation for Fundamental Research on Matter FOM, the Dutch Technology Foundation STW, and an ERC grant.
\end{acknowledgments}

\end{document}